 \documentstyle[aps,prb,epsfig,amsmath,amssymb,floats]{revtex}

 \def \v#1{\mbox{\boldmath$#1$}}
 \def \braket#1{\langle #1 \rangle}

\begin{document}

\twocolumn[\hsize\textwidth\columnwidth\hsize\csname@twocolumnfalse\endcsname

 \title{First-principles study of adhesion at Cu/SiO$_2$ interfaces}

 \author{Kazutaka Nagao$^{1}$\cite{add_nagao}, J. B. Neaton$^2$, and N. W. Ashcroft$^{1}$}
 \address{
              $^{1}$Cornell Center for Materials Research, and 
                    Laboratory of Atomic and Solid State Physics,
                    Cornell University, 
                    Clark Hall, Ithaca, New York 14853-2501\\ 
              $^{2}$Department of Physics and Astronomy, Rutgers University,
                    Piscataway, NJ  08854-8019 
             }
 \date{April 21, 2003}

 \draft
 \maketitle

 \begin{abstract}

 The structural, electronic, and adhesive properties of Cu/SiO$_2$ 
 interfaces are investigated using first-principles density-functional 
 theory within the local density approximation.
 Interfaces between fcc Cu and $\alpha$-cristobalite(001) surfaces 
 with different surface stoichiometries are considered. 
 Interfacial properties are found to be sensitive to the choice 
 of the termination, and the oxygen density at the substrate
 surface is the most important factor influencing the strength of adhesion.
 For oxygen-rich interfaces, the O atoms at the interface
 substantially rearrange after the deposition of Cu layers,
 suggesting the formation of Cu-O bonds.
 Significant hybridization between Cu$-d$ and O$-p$ states is evident in
 site-projected density of states at the interface.
 As oxygen is systematically removed from the interface, less rearrangement
 is observed, implying weaker adhesion.
 Computed adhesion energies for each of the interfaces are found to
 reflect these observed structural and bonding trends,
 leading to the largest adhesion energy in the oxygen rich cases.
 The adhesion energy is also calculated between Cu and SiO$_2$ substrates
 terminated with hydroxyl groups, and adhesion of Cu to these substrates is 
 found to be considerably reduced.
 This work supports the notion that Cu films can adhere well to hydroxyl-free
 SiO$_2$ substrates should oxygen be present in sufficient amounts at the interface.
 \end{abstract}


 \narrowtext

 ]

 \narrowtext

\section{\label{sec_intro} Introduction}

 Adhesion of thin metal films
 to glass substrates have continued to be the subject of intense study for
 many years because of its importance to the large-area electronics industry.
 Copper in particular, a noble metal with a high bulk thermal and electrical conductivities,
 and apparently low electromigration rate,\cite{mur1} is an excellent candidate for 
 interconnects in integrated circuitry on dielectric substrates, and thus the formation of a 
 strong and highly reliable interface between copper and glass has been an issue of primary importance.
 Unfortunately, Cu films are often reported to bind rather
 poorly to oxide-based glass 
 substrates,\cite{6ben,matt,bag1,6lafon2,kriese,ohmi91,6unk,ding1,ding2,shep,wang,ma02,liu,hu} 
 and films that do adhere at room temperature can show diminished adhesivity after 
 thermal cycling,\cite{6unk,wang} an effect often attributed to differences
 of thermal expansion between Cu and the oxide substrate.\cite{6unk}
 In many studies, poor bonding is frequently explained by noting that since Cu does 
 not reduce SiO$_2$, a graded oxide layer facilitating the adhesion 
 is unable to form. 
 Chemically, copper possesses a half-empty $s$ shell
 and a reasonably tightly-bound and filled
 $d$-shell, and thus is less reactive than aluminum and many transition 
 metal films,\cite{6ben} which oxidize well and are observed to adhere more reliably.
 Thus it is not surprising that adhesion is often improved experimentally by applying
 transition metal intermediary layers\cite{kriese,6unk} prior to Cu film deposition, 
 or alloying the Cu films with small amounts
 of Mg or Al.\cite{bag1,6lafon2,ding1,ding2,shep,wang}

 Adhesion of copper films to oxide glasses is, of course, expected to be
 highly sensitive to the temperature, oxygen partial pressure,
 and the condition of the glass surface prior to deposition.
 A common feature of many studies of Cu films on glass is that they were performed under 
 conditions amenable to surface passivation either before or during deposition
 of the copper film.
 For example, Cu films sputtered under high purity conditions 
 in a vacuum {\it do} adhere to glass
 substrates.\cite{6lafon2,ohmi91,6unk}
 In particular, Ohmi {\it et al.}\cite{ohmi91} demonstrated
 robust adhesion of copper to glass after removing hydroxyl (OH) groups 
 from the interface {\it in situ} just prior to deposition. 
 This suggests that dangling bonds at moisture-free glass surfaces
 would be readily saturated by Cu-O bonds.

 Glasses important to industry can be chemically complex, often containing
 a substantial amount (perhaps $\sim 30$\%) of aluminum, boron, and
 other alkaline earth elements.
 Their primary component, however, is almost always SiO$_2$,
 which by itself forms one of the simplest glasses, amorphous silica
 (a-SiO$_2$). In the interest of simplifying the chemistry, 
 we shall restrict our focus to chemically-pure SiO$_2$ glasses for 
 the duration of this study.
 Amorphous silica has no long range order yet remarkably,
 on length scales comparable to a Si-O bond, it is nearly perfectly ordered:
 the bond-length does not vary appreciably from 1.61~{\AA}; each Si atom 
 is tetrahedrally coordinated with O atoms, and the bonds are primarily of 
 a covalent nature.  
 At length scales between 5~{\AA} and 8~{\AA}, minimal variations in O-Si-O and Si-O-Si bond 
 angles result in an appreciable degree of intermediate range orientational disorder.  
 Previous work using molecular dynamics has been performed on both bulk vitreous
 silica and silica surfaces.\cite{garofalini1,garofalini2,garofalini3,garofalini4,car} 
 Much of it was based on semi-empirical potentials,\cite{garofalini1,garofalini2,garofalini3,garofalini4}
 which though able to simulate systems large enough to correctly capture the effects
 of the intermediate range orientational disorder, are expected to be far less 
 accurate when local conditions deviate from those in the bulk, such as they do at 
 surfaces or interfaces.

 Since microscopically adhesion is related to the strength of
 the electronic bond between atoms at the interface,
 the {\it local} electronic structure at the interface will play an important role
 in understanding the reactivity of the metal with the oxide substrate.
 A first principles treatment would therefore be most appropriate
 for examining the nature of local bonding at an ideal interface between
 the two systems, and indeed this approach has proved insightful in the past
 for studying adhesion between metal films and insulating substrates.\cite{dudiy01,finnis,hoek98,siegel02}
 Previous first-principles work on the $\alpha-$quartz surface\cite{car}
 observed significant changes in bonding near the surface; to our knowledge,
 however, the effects of metallic overlayers on SiO$_2$ surface
 geometries and electronic properties have yet to be examined from
 first principles.
 In this article, we assess the degree to which the 
 chemical bonding at atomically-sharp interfaces reflects the empirical trends in the adhesion of
 metallic contacts to glass substrates through study of a simplified system, 
 Cu monolayers on $\alpha-$cristobalite, a crystalline SiO$_{2}$ polymorph
 which although ordered does possess a density quite close to that of a-SiO$_{2}$.\cite{downs94}
 Adhesion is studied as a function of oxygen surface coverage from first principles, 
 using density functional theory within the local density approximation (LDA).\cite{6hks}
 Our methods allow us to elucidate the role of local chemistry in binding the film 
 to glass, but require us to neglect its inherent long-range
 disorder and also, to a limited extent, its intermediate-range orientational disorder.  
 Since we are chiefly interested in the local bonding properties, which
 are expected to be the same for both amorphous and crystalline silica, the
 loss of disorder is acceptable.
 We summarize the method and approximations used here in Sec.~II;
 in Sec.~III we describe the relaxed 1$\times$1 $\alpha$-cristobalite(001)
 surfaces considered here, which include both stoichiometric and non-stoichiometric surface 
 terminations, chosen so as to reflect different possible oxygen coverages that may result
 from different deposition conditions.
 In Sec.~IV, we detail results of atomic relaxation of 
 Cu/SiO$_{2}$ interfaces, where the adhesion is
 found to be critically sensitive to oxygen density at the substrate surface,
 and where relatively strong bonding is reported for oxygen-rich surfaces.
 The effect of hydroxyl groups as a surface passivator is 
 also briefly considered by calculating adhesion energies
 between copper monolayers and a hydroxyl-terminated SiO$_2$ surface.
 After discussing our results and their implications, we provide concluding
 remarks in Sec.~V.

%-------------------------------------------
 \section{\label{sec_method} Methodology}
%-------------------------------------------

\subsection{Computational details}

 Our first-principles density-functional calculations are carried out within the LDA,
 using the correlation energy of Ceperley and Alder,\cite{ca80} as implemented
 within the Vienna {\it ab initio} Simulation Program\cite{kresse93,kresse96a,kresse96b}
 (VASP).
 We use a plane-wave basis set with a 29 Ry cut-off,
 and ultrasoft pseudopotentials\cite{van90,kresse94} for Cu and H,
 projector augmented-wave potentials\cite{blockl94,kresse99} for O,
 and a norm-conserving pseudopotential\cite{Rappe90} for Si. 
 The Monkhorst-Pack $\v{k}-$point mesh is chosen so that 
 $n_{k}=3\times 3\times 1$ for the $5{\rm~{\AA}}\times 5{\rm~{\AA}}\times 30{\rm~{\AA}}$
 supercell which is used in this work.
 Also, we make use of a Fermi-distribution smearing with
 the temperature of $k_{B}T\sim 0.2$ eV to facilitate rapid
 convergence.
 These computational conditions provide good convergence of the
 structure, density of states, and energy differences given here, 
 and are used throughout this work unless otherwise stated.

 It should be mentioned that the
 $\v{k}-$point sampling discussed above is not entirely sufficient for study of 
 the bulk properties of copper.
 In fact, the total energy of bulk copper obtained
 with this sampling is converged to about 0.02 eV per Cu atom.
 However, a sparser mesh turns out to be adequate for this study
 since energy differences are converged.
 To demonstrate, we increased our sampling to $n_{k}=4\times 4\times 1$ for
 the supercell mentioned above,
 recalculated total energies and forces, and found that the resulting difference
 in adhesion energy is less than 4\% even in the largest case.

\subsection{Bulk properties of constituent systems}

%---------------------------------------------------------------------------
 \begin{table}
 \caption{\label{tab_bulk}
          Lattice constants of Cu and $\alpha$-cristobalite
          observed in experiment and LDA calculation.\protect\cite{ref_cutoff}
          Wyckoff positions, two Si-O bond lengths, and four 
          O-Si-O bond angles are also shown for $\alpha$-cristobalite; the Si Wyckoff
          position (4a) is shown as $(u,u,0)$ and
          the O Wyckoff position (8b) as $(x,y,z)$.
          The experimental data are taken from Refs.~\protect\onlinecite{cu_exp}
          and \protect\onlinecite{acr_exp} for copper and $\alpha$-cristobalite,
          respectively.
          }
% \begin{ruledtabular}
 \begin{tabular}{rrll} 
                       &                   &  Experiment  &  This work (LDA)\\ \hline 
  Cu \hspace{5mm}      &        $a$ (\AA)  &  3.615       &   3.532         \\
                       &$\sqrt{2}a$ (\AA)  &  5.112       &   4.995         \\ \\
  $\alpha$-cristobalite&        $a$ (\AA)  &  4.972       &   4.975         \\
                       &        $c$ (\AA)  &  6.922       &   6.907         \\ 
                       &        $u$ (Si)   &  0.3003      &   0.3005        \\
                       &        $x$ (O)    &  0.2392      &   0.2389        \\
                       &        $y$ (O)    &  0.1044      &   0.1058        \\
                       &        $z$ (O)    &  0.1787      &   0.1801        \\
                       &       Si-O (\AA)  &  1.6026      &   1.6056        \\
                       &       Si-O (\AA)  &  1.6034      &   1.6062        \\
                       &O-Si-O ($^{\circ}$)&  109.0       &   108.3         \\
                       &O-Si-O ($^{\circ}$)&  110.0       &   109.9         \\
                       &O-Si-O ($^{\circ}$)&  108.2       &   108.8         \\
                       &O-Si-O ($^{\circ}$)&  111.4       &   111.7         \\
 \end{tabular}
% \end{ruledtabular}
 \end{table}
%-----------------------------------------------------------------------

 Copper crystallizes in the fcc structure under normal
 conditions and its lattice constant is 3.615~{\AA} at 291 K.\cite{cu_exp}
 The LDA underestimates this lattice constant as is well known, and we obtain
 3.532~{\AA} for the equilibrium value as shown in Table~\ref{tab_bulk},
 in agreement with a previous LDA study.\cite{khein95}
 At equilibrium we find $a$ = 4.975~{\AA} and $c$ = 6.907~{\AA} for 
 the tetragonal cell of bulk $\alpha$-cristobalite (the $P4_{1}2_{1}2$ structure),
 in excellent agreement with experiment\cite{acr_exp} and previous 
 LDA calculations.\cite{teter95,hafner99} Further, the 
 computed Si-O bond length ($\sim1.606$~{\AA}) is
 remarkably close to experiment ($\sim 1.603$~{\AA}), and the calculated
 O-Si-O bond angle is near the tetrahedral ideal value,
 also in-line with measurement.\cite{acr_exp}
 Conveniently, the choice of $\alpha$-cristobalite 
 circumvents spurious lattice mismatch when considering the interfaces
 with Cu overlayers: Compare the lattice constant $a$ of $\alpha-$cristobalite
 with $\sqrt{2}a$ of Cu in Table~\ref{tab_bulk}.

%----------------------------------------------------------------
\section{\label{sec_substrate}
         (001) surfaces of $\alpha-$cristobalite S$\v{\rm i}$O$_{2}$}
%----------------------------------------------------------------

 As a starting point, we consider an $\alpha-$cristobalite
 slab consisting of 5 SiO$_2$ layers, the thickness of which is $\sim$8.3~{\AA}.
 Translational symmetry is removed along the $c$-axis,
 defining two different (001) surfaces perpendicular to this axis.
 We embed this $\alpha$-cristobalite slab,
 along with about 20~{\AA} of vacuum, within a supercell having dimensions
 of 5~{\AA}$\times$5~{\AA}$\times$30~{\AA}. 
 Oxygen number at the top of our SiO$_{2}$ slab is then changed
 to mimic different substrate terminations.
 Two nonbridging oxygen atoms terminate the bottom of the slab,
 and two additional H atoms are attached to the oxygens to remove dangling bonds.
 Forces on the atoms are computed using the Hellmann-Feynman theorem,
 and the positions of atoms are then updated until the total energy reaches
 a minimum.
 The oxygen atoms at the bottom surface, terminated
 by hydrogens, are fixed at their bulk positions in order to reduce size effects 
 owing to the finite thickness of the SiO$_{2}$ slab, though the positions of each of 
 the hydrogen atoms are permitted to relax.
 Previous first-principles calculations of 
 $\alpha$-quartz surfaces\cite{car} and model Si/SiO$_2$ interfaces\cite{nma} 
 have indicated that about 5~{\AA} away from the interface, the local structural and 
 electronic properties of the slab are bulk-like, and thus 
 the size of our slab is sufficient to approximate both surface 
 and bulk features.

%---------------------------------------------------------------------------
\begin{figure}
 \begin{center}
   \includegraphics[height=0.18\textheight,width=0.47\textwidth,clip]{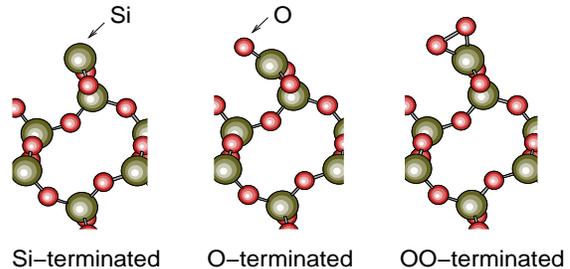}
 \end{center}
  \caption{\label{fig1} Relaxed structures of
                        the $\alpha$-cristobalite(001) surfaces.}
\end{figure}
%---------------------------------------------------------------------------

 Figure~\ref{fig1} presents views of the relaxed $\alpha$-cristobalite(001) surfaces.
 Supercells with slabs having three different surface terminations are considered in this work,
 which we label as Si-, O-, and OO-terminated.
 The Si-terminated slab contains five SiO$_2$ units together with two bottom hydrogens;
 the O-terminated slab is identical to the Si-terminated except that the dangling surface
 oxygen atom is added; the OO-terminated slab is again a modification of
 the Si-terminated slab in which two oxygens are added to the surface.
 Bond lengths and angles are not expected to remain bulk-like
 at the surface, where the choice of termination can leave
 non-bridging oxygens with dangling bonds at the surface (e.g., 
 in the OO-terminated cases).
 Indeed, we find that relaxation significantly changes the Si-O bond lengths
 within two atomic layers from the surface for all terminations considered,
 and the nature of these changes is strongly dependent on the termination. 
 In the Si-terminated case, the additional charge carried by the dangling
 Si bonds forces the Si surface atom above the plane of the surface;
 as a result the Si-O bonds supporting it from below elongate slightly
 to 1.64~{\AA}.
 In the O-terminated case, the surface O atom, bonded to only one other Si, 
 relaxes downward {\it into} the surface, and its bond length decreases by more 
 than 5\% from 1.6~{\AA} to 1.5~{\AA}.
 In the OO-terminated case, in some contrast to the Si- and O-terminated slabs,
 the Si-O bond lengths remain essentially the same as bulk.
 But the distance between
 O neighbors on the surface is also 1.6~{\AA}, which is very small
 compared with typical bulk O second neighbor distances of 2.7~{\AA}.
 Thus O atoms may evidently saturate dangling bonds
 through bonding with neighboring O atoms at this surface.

 As expected, the total cohesive energies of the supercells decrease
 substantially with each additional O atom:
 adding one oxygen atom or two oxygen atoms lowers the cohesive energy
 by $\sim$7.1 eV or $\sim$11.3 eV per surface Si, respectively.
 Since the energy required to break the O$_2$ bond is nearly 5 eV per molecule,
 this energy comparison suggests that the OO-terminated surface may be the most
 stable of the three terminations, and the O-terminated surface the second.
 Evidently our first-principles calculations are consistent
 with the intuition that glass surfaces should be oxygen-rich.

%----------------------------------
 \section{\label{sec_IF} C$\v{\rm u}$/S$\v{\rm i}$O$_2$ interfaces: results and discussion}
%----------------------------------

\subsection{\label{sec_IF_A} Interfacial structure}

 We now add five Cu layers to each of the fully-relaxed (001) $\alpha$-cristobalite surfaces, 
 obtained as discussed above in Sec.~III. The Cu layers are initially positioned on the surface
 so that a Cu atom in the lowest monolayer lies on the former symmetry axis of
 the $\alpha$-cristobalite slab.
 The length of the supercell lattice vector normal to the surface (the $c$-axis) 
 is fixed to 30~{\AA} as in Sec.~III,
 a value we find to be large enough to accommodate up to
 five additional Cu layers while keeping the interactions between 
 the surfaces in neighboring supercells minimal.
 Each monolayer in our supercell contains four Cu atoms,
 and before relaxation each Cu atom has four intralayer neighbors
 at a distance of 2.5~{\AA}.
 Fortunately, the lattice parameter of the $1\times 1$ surface of $\alpha$-cristobalite
 is well matched to that of the $\sqrt{2}\times\sqrt{2}$ surface of
 copper as shown in Table~\ref{tab_bulk} (the mismatch is less than {1\%}).
 The small lattice mismatch is artifact of our approximate treatment of
 amorphous silica as crystalline.  In reality, stress induced by lattice-matching 
 is relieved through formation of defects and/or dislocations. (In the case of
 a truly amorphous substrate, stress may also be overcome through surface 
 reconstruction, a complexity we neglect in the present work.)
%

%---------------------------------------------------------------------------
\begin{figure}
  \vspace*{-15mm}
  \begin{center}
   \includegraphics[height=0.25\textheight,width=0.50\textwidth,clip]{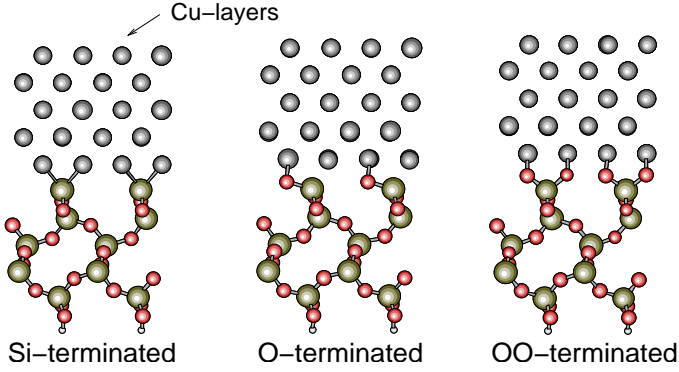}
  \end{center}
  \caption{\label{fig2} Relaxed structures of the Cu/$\alpha$-cristobalite(001)
           interfaces.}
\end{figure}
%---------------------------------------------------------------------------

 The positions of all atoms in the supercell
 are again relaxed except for the bottom O atoms, and
 Fig.~\ref{fig2} shows the optimized structures of the
 interfaces for each termination.
 The most significant reconstruction at the interface
 is observed in the OO-terminated case.
 The O-O bond observed at the free $\alpha$-cristobalite surface
 in Fig.~\ref{fig1} is broken by the deposition of copper,
 and the O-Si-O angle is changed from 59$^{\circ}$ to 104$^{\circ}$ 
 with the Si-O bond lengths kept almost constant. 
 This large rearrangement would indicate bond formation between 
 interfacial O and Cu atoms, suggesting good chemical adhesion 
 between the oxide substrate and the Cu film.
 Each interfacial O atom has two Cu neighbors,
 and the Cu-O bond lengths are computed to be $\sim 1.9$~{\AA},
 strikingly similar to Cu-O distances found in cuprates,\cite{cup} where each
 copper atom is coordinated by four oxygens and considered to be in
 a Cu$^{2+}$ state.
 Similar Si-O and Cu-O bond lengths are observed at the interface between O-terminated 
 surfaces and Cu monolayers, although the Si-O-Cu angles are smaller than in 
 the OO-terminated case by $\sim 15^\circ$, as can be seen in Fig.~\ref{fig2}.
 The magnitude of the reconstruction at the O-terminated/Cu interface
 is less than that at the OO-terminated/Cu interface.

 The interface between the Si-terminated substrate and Cu is completely different 
 from either of the oxygen terminated interfaces.
 In this case, the interfacial Si atom possesses four Cu neighbors, each
 with bond length $\sim 2.4$~{\AA}, and there is negligible atomic rearrangement,
 implying more metallic-like bonding.

\subsection{Electronic properties}

%---------------------------------------------------------------------------
\begin{figure*}	
  \begin{center}
   \includegraphics[height=0.3\textheight,width=0.7\textwidth,clip]{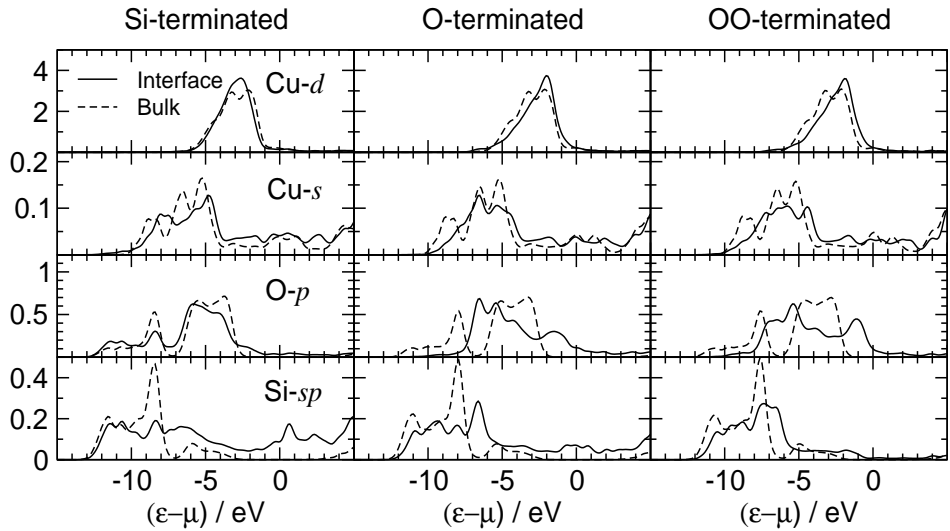}
  \end{center}
  \caption{\label{fig3} Local density of states (LDOS) projected by site
           and angular momentum.
           The solid lines show the LDOS of atoms at (or near) the interfaces, 
           and the dashed lines show those of atoms in the middle
           of the slabs (the LDOS of an approximate bulk material).
           When computing the LDOS, a gaussian smearing
           method is used with $\sigma=0.2$ eV.
           The sphere sizes of each atom are taken as 1.11~{\AA} for Si,
           0.73~{\AA} for O, and 1.38~{\AA} for Cu.  }
\end{figure*}
%---------------------------------------------------------------------------
%---------------------------------------------------------------------------
\begin{figure}
%  \vspace*{-15mm}
  \hspace*{-5mm}
  \begin{center}
  \includegraphics[height=0.35\textheight,width=0.4\textwidth,clip]{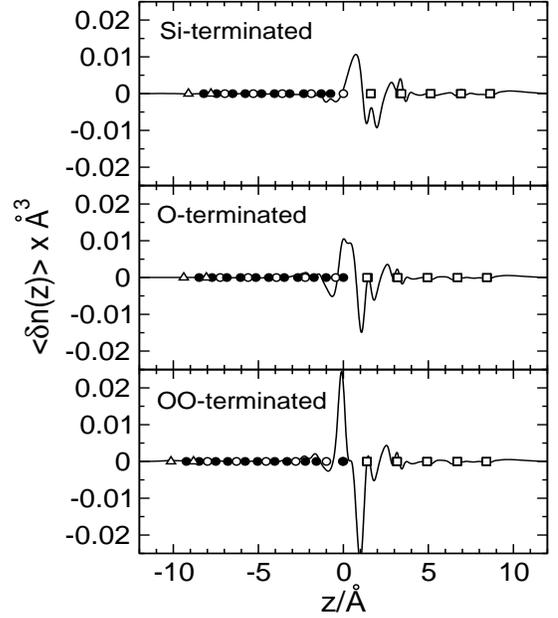}
  \end{center}
%  \vspace*{-18mm}
  \caption{\label{fig4} Change in electron density 
           $\braket{\delta n(z)}$ (see text) as
           a function of the depth $z$ from the substrate surface.
           The positions of each atom are designated by open circles for Si,
           solid circles for O,
           open squares for Cu, and open triangles for H.}
\end{figure}
%---------------------------------------------------------------------------

 To investigate further the bonding properties of each interface
 discussed above, we compute the site-projected local density of states (LDOS) for
 atoms near the interfaces, and compare them with those obtained 
 for bulk-like atoms deeper inside the slabs; the LDOS are
 shown together for each interface in Fig.~\ref{fig3}.
 In the OO-terminated case,
 significant hybridization is observed between
 Cu$-d$ and O$-p$ states just below the Fermi level.
 A key feature is the slight bump in the O$-p$ LDOS, lying about 1 eV below the Fermi energy,
 which becomes progressively smaller with decreasing O content at the interface, and nearly
 vanishes in the Si-terminated case.
 Evidently strong hybridization between Cu$-d$ and O$-p$ states is responsible
 for the significant reconstruction at the OO-interface.
 It is worth pointing out here that
 as the number of O atoms at the interface is decreased,
 the Si$-sp$ LDOS around the Fermi energy increases, 
 that is the interfacial bonding takes on a more metallic 
 character.
 This behavior results from the hybridization
 of Si$-sp$ states with the itinerant Cu$-s$ states around
 the Fermi energy.
 In addition, we observe the bandgap computed within the LDA 
 (an underestimate of the true bandgap) opens and gradually approaches
 its bulk LDA value for LDOS in the middle of the SiO$_2$ slab, 
 confirmation that our slab exhibits approximate bulk behavior
 away from the interface.

 To examine the degree to which changes in structural and bonding properties are 
 confined to the interface, and to observe the mixing between
 the electronic states of the SiO$_{2}$ substrate and the Cu layers in more detail,
 we calculate the density difference from the superposition,
 namely,
 \begin{eqnarray}
   \braket{\delta n(z)} &=& \braket{n_{\rm IF}(z)}
                           -\left[
                                   \braket{n_{\rm SiO_{2}}(z)}
                                  +\braket{n_{\rm Cu}(z)}
                            \right], \nonumber
 \end{eqnarray}
 where $\braket{n_{\rm IF}(z)}$, $\braket{n_{\rm SiO_{2}}(z)}$,
 and $\braket{n_{\rm Cu}(z)}$
 are the densities of the Cu/SiO$_{2}$ interface, 
 the SiO$_{2}$ substrate, and the Cu layers, respectively, averaged over the $xy$ plane
 (parallel to the interface).
 To obtain $\braket{n_{\rm SiO_{2}}(z)}$ ($\braket{n_{\rm Cu}(z)}$),
 we simply remove the Cu layers (SiO$_{2}$ substrate) from
 the fully relaxed interface in the supercell,
 and then recalculate the electronic structure self-consistently keeping the 
 atoms fixed.
 The quantity $\braket{\delta n(z)}$ thus indicates
 the change in electron density resulting from chemical bonding between 
 the Cu layers and the SiO$_{2}$ substrate. (To simplify the analysis, we
 are neglecting additional changes stemming from the atomic
 rearrangements at the free surfaces of SiO$_{2}$ and Cu.)
 The results of calculations of the $\braket{\delta n(z)}$
 appear in Fig.~\ref{fig4}.
 The $\braket{\delta n(z)}$  in the OO-terminated case most prominently
 deviates from zero around the interface, reflecting significant
 charge transfer between the two slabs.
 The charge transfer in the O- and Si-terminated cases is comparatively
 much smaller.
 In particular, in the Si-terminated case, the depletion of the density
 in the vicinity of the lowest Cu layer is not so drastic,
 suggesting that the localized Cu$-d$ states have less influence on
 the bonding around the interface.

\subsection{Quantitative analysis of adhesion}

%---------------------------------------------------------------------------
 \begin{table}
 \caption{\label{tab_wsep}
           Adhesion energy (ideal work of separation)
           $W$ of the Cu/$\alpha$-cristobalite(001) interfaces.
          }
% \begin{ruledtabular}
 \begin{tabular}{rc}
                &  $W$ (J/m$^{2}$) \\ \hline
  Si-terminated &  1.406 \\ 
  O-terminated  &  1.555 \\ 
  OO-terminated &  3.805 \\ 
 \end{tabular}
% \end{ruledtabular}
 \end{table}
%-----------------------------------------------------------------------

 To assess the adhesive strength of the interfaces, we have calculated 
 the ideal work of separation, or {\it adhesion energy}, per unit area $W$ from
 \begin{eqnarray}
   W &\equiv& (E_{\rm SiO_{2}}+E_{\rm Cu}-E_{\rm IF})/A,  \nonumber
 \end{eqnarray}
 where $E_{\rm SiO_{2}}$, $E_{\rm Cu}$, and $E_{\rm IF}$ are the energies
 of the isolated SiO$_{2}$ substrate, Cu layers, and
 Cu/SiO$_{2}$ interface in the supercell, respectively, and $A$ is the area.
 Physically, the adhesion energy $W$ is the work per unit area required
 to separate the interface into the Cu layers and the SiO$_{2}$ substrate within
 a microcanonical process, and it can be considered a 
 measure of the strength of the adhesion.
 For the purposes of comparison, all energies are calculated using 
 supercells of the same size (5~{\AA}$\times$5~{\AA}$\times$30~{\AA}) 
 independent of whether it contains Cu layers, any of the SiO$_{2}$ substrates, or
 Cu/SiO$_{2}$ interfaces.
 Table~\ref{tab_wsep} lists the adhesion energy for each
 case.
 The adhesion energy in the OO-terminated case turns out to be
 much larger than those in the O- and Si-terminated cases, and comparable
 to values previously computed for other metal-dielectric interfaces,
 such as Co/TiC (Ref.~\onlinecite{dudiy01}) and Nb/Al$_2$O$_3$ (Ref.~\onlinecite{finnis}).
 We note that the computed energies for the OO-terminated case, $\sim$~4~J/m$^{2}$,
 are consistent with values obtained experimentally by Kriese {\it et al.}\cite{kriese}
 for 100~nm thick films of Cu on SiO$_2$ using an indentation technique.
 In summary, the magnitudes of adhesion energies computed here for each of 
 the three different interfaces reflect the trends witnessed above in their 
 structural and bonding properties.

\subsubsection{Local atomic structure}

%---------------------------------------------------------------------------
\begin{figure}
  \begin{center}
   \includegraphics[height=0.28\textheight,width=0.4\textwidth,clip]{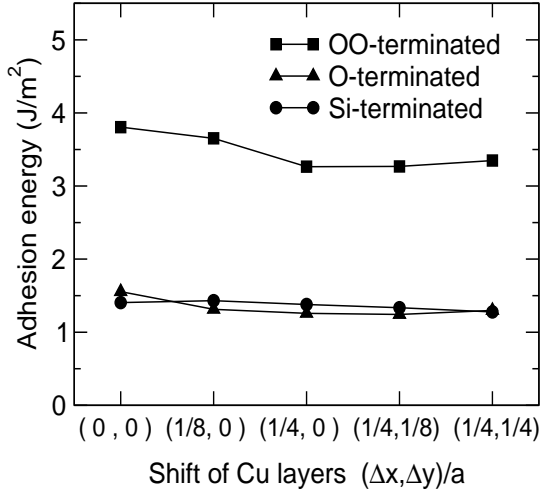}
  \end{center}
  \caption{\label{fig5}
           Adhesion energy as a function of 
           the position of Cu layers in the $xy$ plane
           calculated for five different initial positions 
           and for each of the three terminations. Each point is obtained
           by translating Cu layers [initially positioned on the substrate
           as described in Sec.~\ref{sec_IF_A} $(\Delta x=0,\Delta y=0)$]
           in the $xy$ plane by $(\Delta x,\Delta y)$ and then fully relaxed.
           The abscissa (which quantifies the magnitude of
           the shift of Cu layers) is scaled by 
           the supercell lattice parameter $a$=5~{\AA}.
           }
\end{figure}
%---------------------------------------------------------------------------

 At this point, it is meaningful to investigate further a drawback
 of using crystalline $\alpha-$cristobalite to model the 
 amorphous substrate.
 When Cu layers are placed on an amorphous SiO$_{2}$ substrate,
 different local atomic configurations are possible
 at the interfaces because of the non-periodicity of glass.
 A principal simplification of our model is the imposition of translational
 symmetry: we are unable to directly assess the difference between
 the adhesive properties of crystalline substrates and 
 those of amorphous substrates.
 We are, however, able to investigate the sensitivity of our calculated adhesion energies
 to changes in local atomic structure. As a first step, we compute
 the influence of the position of Cu layers, relative to the substrate, 
 on the adhesion energy by translating the Cu layers in the $xy$ plane.
%

%---------------------------------------------------------------------------
\begin{figure}
  \vspace*{-15mm}
  \begin{center}
  \includegraphics[height=0.3\textheight,width=0.50\textwidth,clip]{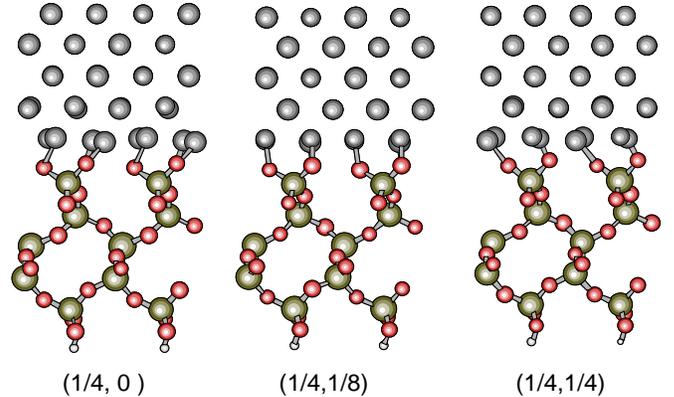}
  \end{center}
  \caption{\label{fig6}
           Structures at the OO-terminated interfaces obtained
           as a function of initial positions of Cu layers with
           respect to the substrate. The figures within brackets denote the shift
           of the position of Cu layers within the $xy$ plane
           (see the caption of Fig.~\protect\ref{fig5})
           in unit of 5~{\AA} (the in-plane supercell lattice parameter).
           }
\end{figure}
%---------------------------------------------------------------------------
%
 As shown in Fig~\ref{fig5}, the adhesion energy turns out rather
 insensitive to the initial position of the Cu layers
 so long as the termination type remains unchanged. 
 This result suggests that the adhesive properties realized
 in these $\alpha-$cristobalite substrates may well be carried over, 
 to some extent, to those of glass substrates.
 The insensitivity of adhesion energy to the position of Cu layers
 does {\it not} imply that the detailed structure at the interface is less
 important for adhesion.
 In fact, as shown in Fig.~\ref{fig6} for the OO-terminated case,
 atoms at the interfaces do move significantly relative to
 the unshifted case.
 Since the surface atoms have a greater freedom to choose their position
 than those in bulk, the adhesion energy remains
 rather constant for various initial positions of Cu layers.

\subsubsection{Local oxygen surface coverage}

 Thus far we have investigated the Si-, O-, and OO-terminations
 independently. Depending on the oxygen partial pressure, however,
 these terminations are generally expected to coexist on the substrate
 surface, independent of whether the substrate is crystalline or amorphous.
 To examine the effects of oxygen surface densities {\it intermediate} between
 those of the Si-, O-, and OO-terminated cases,
 we enlarge the supercell by a factor of two along the $y$-axis
 (along the horizontal direction of Figs.~\ref{fig1}, \ref{fig2}, and \ref{fig6}).
 For the 1$\times$2-supercells, three new surfaces, in addition to those
 already considered above, are now possible. These are:
 a (Si,O)-terminated surface which consists of neighboring Si- and O-terminated
 surfaces;
 a (Si,OO)-terminated surface which consists of neighboring Si- and OO-terminated surfaces; and
 a (O,OO)-terminated surface which consists of neighboring O- and OO-terminated surfaces.
 [Note that the (Si,OO)-terminated case possesses the same oxygen density
 as the O-terminated case.]
 We find that the (Si,OO)-terminated substrate has
 energy which is higher than that of the O-terminated surface
 by $\sim 1.4$ eV per Si atom at the surface (per 25~{\AA}$^{2}$). 
 Therefore, we restrict our focus to investigation of the (Si,O)- 
 and (O,OO)-terminated cases, calculating their relaxed structures
 and adhesion energies.

%---------------------------------------------------------------------------
\begin{figure}
  \begin{center}
  \includegraphics[height=0.25\textheight,width=0.4\textwidth,clip]{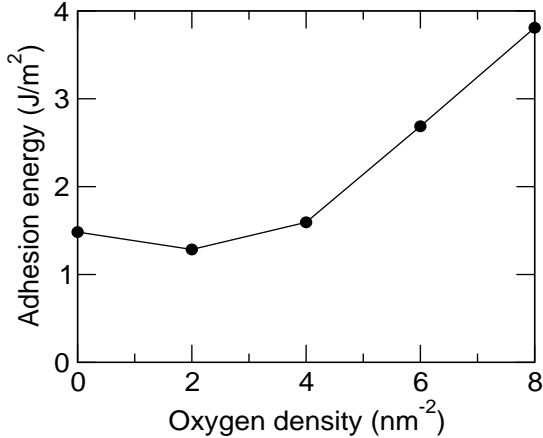}
  \end{center}
  \caption{\label{fig7}
           Adhesion energy as a function of oxygen density
           at the substrate surface.\protect\cite{ref_kpoint}
           The densities $0$, $2$, $4$, $6$, and $8$ nm$^{-2}$
           correspond to the Si-, (Si,O)-, O-,
           (O,OO)-, OO-terminated cases, respectively.
           }
\end{figure}
%---------------------------------------------------------------------------

 Figure~\ref{fig7} depicts the adhesion
 energy as a function of the oxygen density at the interface.
 Remarkably, the adhesion energy is {\it not} 
 a monotonically-increasing function of oxygen density;
 at low oxygen density, increasing the number of O atoms a
 small amount does not necessarily lead to stronger adhesion.
 However, beyond the oxygen density of 4 nm$^{-2}$
 where the OO-termination will begin to appear appreciably,
 the adhesion energy increases, and does so almost linearly.
 This linear dependence implies that each OO surface unit contributes 
 to the adhesion fairly independently, and thus
 the locally OO-terminated regions may not necessarily
 have to form large domains in order to work as an effective ``glue'',
 although it is clear that a critical density of such locally
 OO-terminated regions is required for strong adhesion.

\subsubsection{Hydroxylated surfaces}

%---------------------------------------------------------------------------
\begin{figure}
  \vspace{-15mm}
  \begin{center}
  \includegraphics[height=0.35\textheight,width=0.4\textwidth,clip]{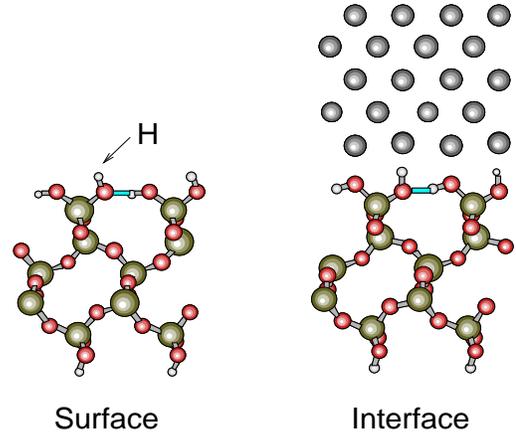}
  \end{center}
  \vspace{-5mm}
  \caption{\label{fig8}
           Relaxed structures of the OHOH-terminated surface and interface.
           }
\end{figure}
%---------------------------------------------------------------------------

 We now address a frequently-cited reason
 for the weak adhesion at Cu/a-SiO$_{2}$ interfaces, namely
 the existence of hydroxyl groups on the SiO$_{2}$ surface.\cite{ohmi91}
 It is well known that hydroxyls are likely to form
 on the SiO$_{2}$ surface under ``wet'' deposition conditions;
 to examine the impact of hydroxyls on adhesion from first-principles, 
 we add two hydrogens to the 1$\times$1 OO-terminated surface, and 
 relax all of the atoms.
 The result is that the cohesive energy of
 OHOH-terminated substrate, in which a Si atom
 is terminated by two hydroxyls, is lower than that
 of OO-terminated substrate by $\sim$ 11.2 eV
 per surface Si atom.
 This energy difference is much larger than the dissociation
 energy of H$_{2}$ molecule ($\sim 4.5$ eV per molecule),
 suggesting the high stability of the OHOH-terminated substrate.
 If we now place
 the Cu layers on top of the hydroxylated surface and relax
 the interface (Fig.~\ref{fig8}), we find that
 the deposition of Cu layers does not significantly affect
 the structure at the OHOH-terminated surface, implying
 little interaction; the hydrogen present at the surface
 leaves the surface neutral and inert.
 This tendency is clearly reflected in the calculated adhesion energy,
 and according to our calculations it is just 0.331 J/m$^{2}$,
 less than one tenth of the adhesion energy observed in the 
 OO-terminated case. 

 However, as shown in Table~\ref{tab_wsep},
 the adhesion energies for all other cases studied here 
 (i.e. the OO-, O-, and Si-terminated cases) seem to 
 be of a magnitude which should produce adhesion, especially
 the oxygen-rich OO-terminated case.
 It is also noteworthy that the OO-terminated substrate,
 which has the largest adhesion energy among the three,
 looks actually the most stable substrate. 
 These facts imply that should dehydroxylation
 of the substrate surfaces be successfully achieved
 (e.g. by annealing or particle bombardment of a growing film surface),
 robust adhesion is entirely possible, at least from the point
 of view of chemical bonding.

\section{\label{sec_concl} Conclusions}

 We have performed a first-principles study of the adhesive
 properties of atomically-sharp Cu/SiO$_{2}$ interfaces.
 As a model a-SiO$_{2}$ substrate,
 we used a crystalline polymorph of SiO$_{2}$, $\alpha-$cristobalite,
 and investigated its Si-, O-, and OO-terminated (001) surfaces in detail.
 For interfaces between Cu and OO-terminated surfaces, a substantial 
 rearrangement of oxygen positions relative to the free surface
 is observed,
 suggesting the formation of strong Cu-O bonds.
 Analysis of the local density of states at the interface
 showed that Cu-O interfacial bonds are composed of Cu$-d$ and O$-p$ states.
 The computed adhesion energy also exhibited a tendency toward
 much stronger adhesion in the OO-terminated interface than
 in the O- and Si-terminated interfaces:
 Substrate surfaces with high oxygen content were found suitable
 for adhesion.

 The adhesion energy is found to be very insensitive
 to the position of the Cu layers if the termination
 type is unchanged, and this observation suggests that 
 $\alpha-$cristobalite may serve as a good starting model
 of a-SiO$_{2}$ substrates.
 The detailed dependence of the adhesion energy on oxygen density
 at the substrate surface has been investigated, and it shows
 that the surfaces where the O- and OO-terminations coexist
 can also lead to relatively strong adhesion.

 The possible existence of hydroxyl groups at the substrate surface
 is thought to be the main cause for the weak adhesion observed
 in experiment.
 However, if these hydroxyls are removed beforehand and 
 the oxygen density at the surface is increased (for example,
 by upping the oxygen partial pressure),
 Cu layers are predicted to adhere reasonably well to SiO$_{2}$.
 In addition to the possibility of direct comparison to future measurements
 on crystalline Cu/$\alpha$-cristobalite(001) interfaces, these results should 
 serve as a starting point for which to study more complicated
 interfacial geometries, benchmarks for future investigations, and
 a baseline for experiments attempting to elucidate complicated
 phenomena affecting adhesion at Cu/glass interfaces.

\section{Acknowledgments}

 We thank S. Baker and M. Backhaus for helpful discussions.
 This work was supported by the National Science Foundation (DMR-9988576),
 and made use of the Cornell Center for Materials Research Shared
 Experimental Facilities, supported through the National Science Foundation
 Materials Research Science and Engineering Centers program (DMR-9632275).


\begin{references}
\bibitem{ref_cutoff} For these calculations, the cut-off energy is increased
                     up to $\sim$38 Ry, and the $\v{k}-$point number
                     is set to $n_{k}=10\times 10\times 10$ for fcc Cu and
                     $n_{k}=4\times 4\times 4$ for $\alpha-$cristobalite.
\bibitem{ref_kpoint} For these calculations, the $\v{k}-$point number is 
                     set to $n_{k}=4\times 2\times 1$ for the
                     $5{\rm~{\AA}}\times 10{\rm~{\AA}}\times 30{\rm~{\AA}}$ supercell.
\end{references}
\end{document}